An index to link scientific productivity with visibility

Ren Zhang

Department of Endocrinology

BIDMC and Harvard Medical School

Boston, MA 02215, USA

E-mail: rzhang@bidmc.harvard.edu

Abstract

I here propose an index that links the number of papers a researcher has published with impact factors (IFs) of the journals that publish these papers. A researcher is said to have an index $z$ if totally $z$ of his/her papers are published in journals with IFs of at least $z/2$. The $z$-index, not meant to evaluate, compare and rank scientists, is a number that hopefully conveniently summarizes the number of publications in journals with high IFs.

The *h*-index, proposed to quantify and compare scientists' research outputs, has been a topic in the whole scientific community (Bornmann & Daniel, 2009), and despite shortcomings, which are not listed here, it has already been widely used. For instance, both of the two largest citation databases, the Web of Science and Scopus, calculate the *h*-index for each published scientist. The original paper of the *h*-index (Hirsch, 2005), having been cited hitherto over 400 times, boosted bibliometric studies and stimulated the proposals of a large number of other evaluative metrics. One important advantage of the *h*-index is that it links two aspects, publications and citations, with one single number, in an extremely simple way.

In the current academic environment, scientists must publish, or they perish. And to survive in this environment, not only must scientists publish a lot (high productivity), they also have to publish in high-impact journals (high visibility). This is more so for postdoctoral fellows trying to gain faculty positions and faculty members trying to gain tenure.

Following the idea of the *h*-index, I here propose an index that links the number of papers a researcher has published with the impact factor (IF) of the journals that publish these papers. Consider that a researcher has published N papers, and then rank the publications in a decreasing order of journal IFs (Fig. 1). A researcher is said to have an index $z$ if totally $z$ of N papers are published in journals with IFs of at least $z/2$. Because IF values tend to be low, the parameter 2 was chosen to increase the discriminatory power. Choosing this parameter was not completely arbitrary, because it bears an important

advantage – being extremely easy to calculate. For instance, a $z$ of 14 means that this researcher has published 14 papers in journals with IFs larger than 7.

I argue that the $z$-index has advantages over some other commonly used single-number indices. The total number of papers measures productivity, but due to lack of information about journal IFs, it can be inflated by a large number of publications in low-impact journals. The sum of IFs, on the other hand, does not have information about paper numbers, and therefore it can be inflated by either one single publication in a high-impact journal, or by a large number of papers in low-impact journals. The $z$-index provides a convenient way to summarize the number of publications in journals with high IFs. The Nobel Prize in Physiology or Medicine of 2009 has been awarded to Drs. Elizabeth Blackburn, Carol Greider and Jack Szostak, for the discovery of telomeres and the telomerase. The $z$-indices for the above three scientists were 38, 28 and 52, respectively, which happened to be considerably high. For instance, Dr. Blackburn has published 38 papers in journals with IFs larger than 19 according to Journal Citation Reports of 2008.

It is necessary to mention some caveats for using this number. As it is well known that papers published in journals with high IFs do not necessarily mean they have high quality. The scientific merit of any work has to be examined carefully in a meaningful context by peer review. Therefore, the $z$-index is not meant to evaluate, compare and rank scientists; it is a number that hopefully provides a convenient way to summarize the number of publications in journals with high IFs.

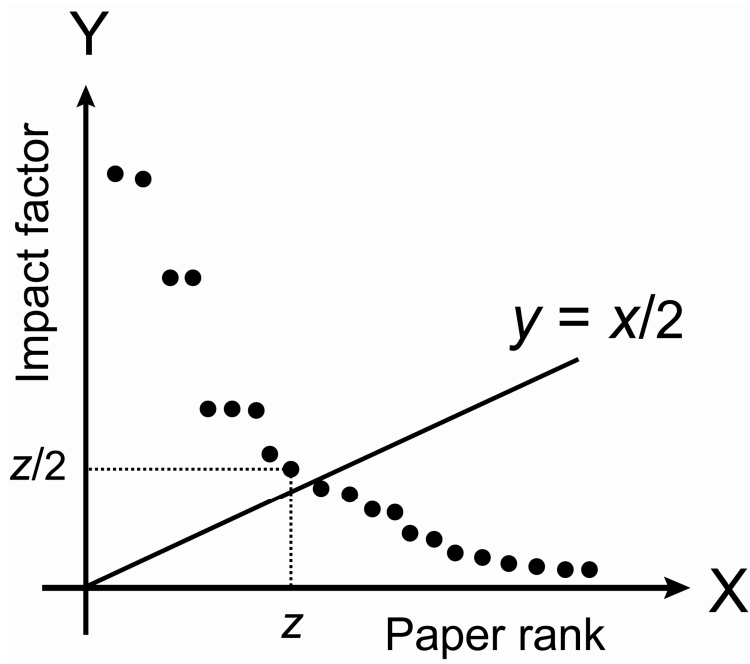

Fig. 1. A schematic diagram showing how the *z*-index is calculated.